\documentclass[a4paper,twocolumn,showpacs,prb]{revtex4}
\usepackage{graphicx}
\usepackage{amssymb}
\usepackage{amsmath,amsfonts,latexsym}

\begin{document}

\title{Effects of Disorder on Coexistence and Competition between
Superconducting and Insulating States}

\author{M. V. Mostovoy$^{1,2}$, F. M. Marchetti$^3$, B. D. Simons$^3$, and
P. B. Littlewood$^3$}

\affiliation{$^1$ Materials Science Center, University of Groningen,
Nijenborgh 4, 9747 AG Groningen, The Netherlands\\ $^2$ Max Planck
Institute for Solid State Research, Heisenbergstrasse 1, 70569
Stuttgart, Germany \\$^3$ Cavendish Laboratory, University of
Cambridge, Madingley Road, Cambridge CB3 0HE, UK}

\date{February 4, 2005}

\begin{abstract}
  We study effects of nonmagnetic impurities on the competition
  between the superconducting and electron-hole pairing. We show that
  disorder can result in coexistence of these two types of ordering in
  a uniform state, even when in clean materials they are mutually
  exclusive.
\end{abstract}








\maketitle

\section{Introduction}
\label{sec:introduction}
At low temperatures, many metals undergo a transition into a state
with a gap in the single-electron excitation spectrum and become
either superconductors or insulators with a periodic modulation of the
electron charge or spin density. The insulating and superconducting
(SC) orders inhibit each other by reducing the fraction of the Fermi
surface available for the gap of the competing phase. The balance
between the two phases is very sensitive to the Fermi surface shape
and can be changed by, e.g., pressure, doping or magnetic field.


Nonetheless, in a surprisingly large number of materials the SC and
insulating states coexist.\cite{Gabovich,Gabovich2} This paper is
focused on the coexistence of superconducting and charge density wave
(CDW) states, observed in, e.g., layered transition metal
dichalcogenides $2H$-NbSe$_2$ and
$2H$-TaS$_2$,\cite{Wilson,Jerome,Berthier} the quasi-one-dimensional
compound NbSe$_3$,\cite{Fuller} tungsten bronzes
A$_x$WO$_3$,\cite{Stanley,Cadwell} and quarter-filled organic
materials.\cite{Mori,Merino} One of the best studied and best
characterized CDW superconductors is the transition metal
dichalcogenide $2H$-NbSe$_2$. At $T_d = 33.5$ K this compound
undergoes a second-order phase transition to an incommensurate CDW
state,\cite{moncton,harper} which is likely driven by the nesting of a
part of the Fermi surface.\cite{straub,Valla} The resistivity,
however, remains metallic-like down to $T_c = 7.2$ K, at which this
material becomes superconducting,\cite{Jerome,Berthier} and the
superconductivity coexists with the CDW modulations.\cite{Du} The
coupling between the CDW and SC order parameters, resulting from the
competition between these two states, was observed in a number of
experiments.  Thus, the suppression of the charge density modulation
by pressure and hydrogen intercalation results in an increase of
$T_c$.\cite{Berthier,Monceau,Murthy} A similar interplay between the
CDW and SC states upon applied pressure and doping is observed in
NbSe$_3$ and tungsten bronzes.\cite{Fuller,Stanley,Cadwell}

In this paper we adopt a rather general, though simplified, viewpoint
on the interplay between the SC and CDW states.  We assume that it
originates from the competition between two different Fermi surface
instabilities: the instability towards the electron pairing, which
gives rise to superconductivity, and the instability towards the
electron-hole (or excitonic) pairing. Here, we focus primarily on the
effects of quenched disorder on this competition. We show that even in
`the worst case scenario', when the two states compete over the whole
Fermi surface and therefore, in absence of disorder, are mutually
exclusive, disorder stabilizes a \emph{uniform} state, in which
superconducting and insulating order parameters coexist. While having
no effect on the superconducting phase, nonmagnetic disorder tends to
close the CDW gap before completely suppressing the corresponding
order parameter. Disorder induces low-energy states by breaking some
of the electron-hole pairs. The released electrons and holes can
subsequently form Cooper pairs, resulting in the coexistence of the
two phases.

While in usual $s$-wave superconductors, non-magnetic impurities have
little effect on the transition temperature,\cite{anderson}
experiments on electron irradiated transition metal dichalcogenides
have shown strong dependence of $T_c$ on the concentration of
defects.\cite{mutka} This was attributed to the interplay between the
SC and CDW orderings: Similarly to effect of
pressure,\cite{Berthier,Fuller} disorder strongly suppresses the CDW
state, which results in the observed increase of the SC critical
temperature. Theoretically, the combined effect of the CDW modulation
and disorder on the pairing instability have been studied in
Ref.~[\onlinecite{Grest}], where an increase of $T_c$ was
found. However, in that paper the amplitude of the CDW modulation was
assumed to be fixed, which is clearly insufficient in view of the
strong suppression of the CDW state by disorder. In this paper we
solve self-consistency equations for both SC and CDW order parameters,
which allows us to study the interplay between these two different
orders and obtain the temperature versus disorder phase diagram of CDW
superconductors.

The remainder of the paper is organized as follows: In
Sec.~\ref{sec:model} we formulate an effective model describing the
interplay between the superconducting and excitonic pairing. The
self-consistency equations for the two order parameters are derived in
Sec.~\ref{sec:equations} and in Sec.~\ref{sec:phd} we analyze the
phase diagram of the model. In Sec.\ref{sec:eisc} we discuss the
electron-hole symmetry underlying the model and its consequences for
the phase diagram. Finally, we conclude in
Sec.~\ref{sec:conclusions}. The details of the derivation of the
effective model can be found in Appendix A.

\section{The model}
\label{sec:model}
\begin{figure}
\begin{center}
\includegraphics[width=0.7\linewidth,angle=0]{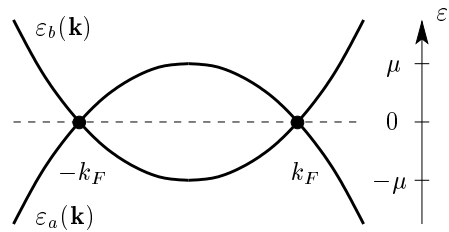}
\end{center}
\caption{\small Schematic picture of the hole-like ($a$) and
    electron-like ($b$) bands.}
\label{fig:bande}
\end{figure}

In the following we consider the microscopic Hamiltonian
\begin{multline}
  \displaystyle \hat{\mathcal{H}} = \int d \mathbf{x}
  \left\{\sum_{\sigma} \sum_{j=a,b}\psi _{j\sigma }^{\dag} \left[
  \varepsilon_{j} (-i \nabla_{\mathbf{x}}) + U(\mathbf{x})\right]
  \psi_{j\sigma} \right. \\
  \left. - g_1 \left(\psi_{a\uparrow}^{\dag}\psi_{a\uparrow}
  \psi_{a\downarrow }^{\dag} \psi_{a\downarrow} +
  \psi_{b\uparrow}^{\dag} \psi_{b\uparrow} \psi_{b\downarrow}^{\dag}
  \psi_{b\downarrow}\right) \right. \\
  \left. + g_2 \sum_{\sigma \sigma'} \psi_{a\sigma}^{\dag}
  \psi_{a\sigma} \psi_{b\sigma'}^{\dag} \psi_{b\sigma'}\right\}
\label{eq:model}
\end{multline}
describing two types of fermions, one with hole-like dispersion
($a$-electrons) and another with electron-like dispersion
($b$-electrons), $\varepsilon_{j=a,b} (\mathbf{k}) = \pm (k_F^2 -
\mathbf{k}^2)/2m$ ($\hbar = 1$), where $\mu = k_{F}^2/2m$ denotes the
chemical potential (see Fig.~\ref{fig:bande}). Here, in comparison
with the models generally used to represent CDW systems, two nested
parts of a single Fermi surface are replaced by two spherical Fermi
surfaces matching at the Fermi wave vector $k_F$. The excitonic
insulator (EI) state is the condensate of pairs formed by
$b$-electrons and $a$-holes (or vice versa) with the zero total
momentum.\cite{Keldysh} It is an analogue of the condensate of
electron-hole pairs with the total momentum $\hbar {\bf Q}$, where
{\bf Q} is a nesting wave vector, appearing in the CDW state.

The disorder potential $U(\mathbf{x})$ encapsulates the effect of
non-magnetic impurities in the system. Here, we assume that the latter
is drawn at random from a Gaussian distribution with zero mean and
variance given by
\begin{equation}
  \langle U(\mathbf{x}) U(\mathbf{y})\rangle = \frac{\Gamma}{2\pi
  \nu_{F}} \delta (\mathbf{x - y}) \; ,
\label{eq:whitenoise}
\end{equation}
where $\Gamma$ is the inverse scattering time and $\nu_{F} = m
k_{F}/2\pi^2$ is the density of states at the Fermi energy. For
simplicity we have assumed the electron and hole effective masses to
be equal.


The interaction term characterized by the coupling strength $g_1$
describes the attraction between electrons of the same type (e.g. due
to the phonon-exchange), while the $g_2$ term describes the (Coulomb)
repulsion between the $a$- and $b$-electrons ($g_1,g_2>0$). The
attraction between electrons favors $s$-wave superconductivity, while
the second interaction leads to an attraction between electrons and
holes and vice versa, favoring the EI state.  Here, we neglect the
inter-band electron transitions due to scattering off impurities and
electron-electron interactions, so that the numbers of the $a$- and
$b$-electrons are separately conserved and fixed by the chemical
potential. Such terms will formally destroy long-range order of the EI
phase, corresponding to the suppression of the long-ranged CDW order,
due to the pinning of the CDW phase by randomly distributed
impurities. However, for the essentially short-length-scale physics we
shall discuss, these effects may be neglected. In the absence of
disorder, the same model~\eqref{eq:model} has been employed to study
the competition between SC and EI states for an arbitrary ratio of
electron and hole densities.\cite{Rusinov} The effect of disorder on
the EI state alone has been considered in the seminal work of
Ref.~[\onlinecite{Zittartz}], where the analogy of the problem to an
$s$-wave superconductor in presence of magnetic
impurities\cite{AbrikosovGorkov} was drawn.

\begin{figure}
\begin{center}
\includegraphics[width=1\linewidth,angle=0]{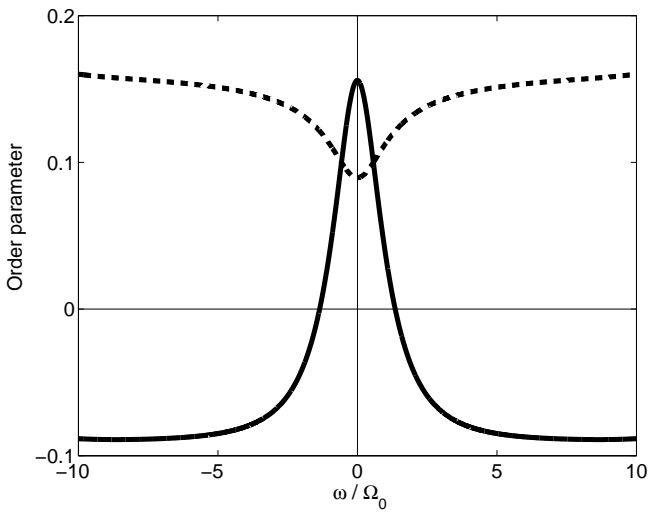}
\end{center}
\caption{\small The dependence of the superconducting (solid line) and
        excitonic (dashed line) order parameters on the Matsubara
        frequency $\omega$ (see appendix~\ref{AppendixA} for
        details).}
\label{fig:opfreq}
\end{figure}

The particular fermion-fermion interactions considered
in~\eqref{eq:model} -- attraction between electrons of the same type
and repulsion between the $a$- and $b$-electrons -- open the
possibility to have simultaneously both superconducting and insulating
instabilities. A more realistic starting point would be a model with
attractive phonon-mediated interactions and Coulomb repulsion between
all types of electrons. However, it is possible to demonstrate that,
since the former are retarded, while the latter is practically
instantaneous, the SC and EI order parameters turn out to have a very
different dependence on the Matsubara frequency $\omega$. This is
clearly shown in Fig.~\ref{fig:opfreq}: The SC order parameter is
large at small frequencies, while at higher values, it decreases in
magnitude and finally changes sign when $\omega$ is of the order of
the phonon frequency $\Omega_0$.\cite{AndersonMorel} By contrast, the
EI parameter is large at high frequencies and has a dip for $|\omega|
< \Omega_0$. In other words, the difference in frequency scales of the
attractive and repulsive interactions allows both instabilities to be
present simultaneously. Furthermore, in the weak coupling limit and
for a weak disorder, i.e. $\Gamma \ll \Omega_0$, the frequency
dependence of the two order parameters can be found separately for
$\omega \sim \Gamma $ and $\omega \gtrsim \Omega_0$. Furthermore, it
can be shown that the self-consistency equations for the order
parameters at $\omega = 0$ coincide with the ones obtained from the
model~\eqref{eq:model}, which therefore can be interpreted as an
effective interacting model. Technical details together with the
frequency dependence of the two order parameters and the explicit
expressions for the coupling constants $g_1$ and $g_2$ in terms of the
Coulomb and electron-phonon couplings can be found in
appendix~\ref{AppendixA}.

%

\section{Order parameters and self-consistency equations}
\label{sec:equations}
Four order parameters describing the SC and EI states can be
introduced by means of the following anomalous averages
\begin{align*}
  \Delta_{1a} &= g_1\langle \psi_{a\uparrow} \psi_{a\downarrow}
  \rangle & \Delta_{1b} &= g_1 \langle \psi_{b\uparrow}
  \psi_{b\downarrow} \rangle \\
  \Delta_{2\uparrow} &= -g_2 \langle \psi_{a\uparrow}^{\dag}
  \psi_{b\uparrow}\rangle & \Delta_{2\downarrow} &= -g_2 \langle
  \psi_{a\downarrow}^{\dag} \psi_{b\downarrow}\rangle \; .
\end{align*}
Since the numbers of the $a$- and $b$-fermions are separately
conserved, for homogeneous states, we use the global gauge
transformation
\begin{align*}
  \psi_{a} &\mapsto e^{i\varphi _{a}} \psi_{a} & \psi_{b} &\mapsto
  e^{i\varphi_{b}} \psi_{b}\; ,
\end{align*}
to make the SC order parameters, $\Delta _{1a}$ and
$\Delta _{1b}$, real and positive. Moreover, as electrons
and holes are characterized by the same dispersion, we can
require, without a loss of generality, that
\begin{equation*}
  \Delta_{1a} = \Delta_{1b} = \Delta_1>0\; .
\end{equation*}
In case of a spin-independent interaction, as in~\eqref{eq:model},
singlet and triplet exciton pairs are degenerate in energy. This gives
rise to a large symmetry class of transformations for the EI order
parameter $\Delta_{2 \sigma}$. In reality, however, this degeneracy is
lifted by Coulomb exchange interactions and inter-band
transitions. Therefore, we will assume the exciton pairs have zero
total spin, i.e. $\Delta_{2\uparrow} = \Delta_{2\downarrow } =
\Delta_2$.

Finally we note that, when $\Delta_1,\Delta_2 \neq 0$ and $\Delta_2$
has an imaginary part, a pairing of electrons of different types,
$\Delta_{1ab} = -g_2 \langle\psi_{a\uparrow} \psi_{b\downarrow}
\rangle = - g_2 \langle\psi_{b\uparrow} \psi_{a\downarrow} \rangle$,
may be present.\cite{Rusinov} However, one can show the energy of the
state with coexisting SC and EI orders to be the lowest for real
$\Delta_2$, in which case $\Delta_{1ab}=0$.

By analogy with the case of magnetic impurities in $s$-wave
superconductors,\cite{AbrikosovGorkov} restricting attention to the
limit in which the disorder potential imposes only a weak perturbation
on the electronic degrees of freedom ($\mu \ll \Gamma$), the mean
field (saddle-point) equations together with the self-consistency
equations for the EI and SC order parameters can be obtained using the
diagrammatic technique. However, we will find it more convenient to
use a path-integral approach. This will also allows us to obtain
straightforwardly an expression for the average free energy.

The quantum partition function, $\mathcal{Z} = \rm{tr} [e^{-\beta
\hat{\mathcal{H}}}]$, where $\beta = 1/T$, can be expressed as a
coherent state path integral over fermionic fields. In order to
facilitate the averaging of the free energy over the disorder
potential~\eqref{eq:whitenoise}, it is convenient engage the replica
trick:\cite{edwards_anderson}
\begin{equation*}
  F = - \frac{1}{\beta} \langle \ln \mathcal{Z}\rangle = -
  \frac{1}{\beta} \lim_{n\to 0} \frac{\langle \mathcal{Z}^n \rangle
  -1}{n} \; .
\end{equation*}
Once replicated, a Hubbard-Stratonovich transformation can be applied
to decouple the interaction terms in the Hamiltonian. As a result, one
obtains:
\begin{multline*}
  \mathcal{Z}^n = \int D (\Psi ,\Psi^\dag) \int D\Delta _1
  D\Delta_2 \\
  \exp \left\{\int_0^{\beta} d\tau \int d\mathbf{x} \left[\Psi^\dag
  \left(\partial_\tau + \hat{H}\right)\Psi + 2 \frac{\Delta_1^2}{g_1}
  + 2 \frac{\Delta_2^2}{g_2}\right]\right\}\; .
\end{multline*}
Here, omitting the replica indices for clarity, the fermion field is
arranged in a Nambu-like spinor, $\Psi^{\rm T} = (\psi_{b\uparrow} ,
\psi_{a\uparrow} , \psi_{b\downarrow}^\dag ,
\psi_{a\downarrow}^\dag)$, in such a way the single quasi-particle
Hamiltonian takes the following form:
\begin{equation}
  \hat{H} = \hat{\xi}_{\hat{\mathbf{p}}} \tau_{3} \sigma_{3} +
  U(\mathbf{x}) \tau_{3} + \Delta_1 \tau_1 + \Delta_2
  \tau_{3}\sigma_1\; ,
\label{Hamiltonian}
\end{equation}
where, $\hat{\xi}_{\hat{\mathbf{p}}} = - \nabla^2_{\mathbf{x}}/2m -
\mu$ and the Pauli matrices $\tau _{c}$ and $\sigma _{c}$ ($c=1,2,3$)
act, respectively, in the particle-hole and the $b,a$ subspace.

The ensemble average over the quenched random potential
distribution~\eqref{eq:whitenoise} induces a time non-local quartic
interactions, $\int d\mathbf{x} (\int_0^\beta d\tau \Psi^\dag \tau_3
\Psi)^2$, which can be decoupled by means of a Hubbard-Stratonovich
transformation with the introduction of a matrix field,
$\Sigma(\mathbf{x})$, local in real space, and carrying replica,
Matsubara ($\omega_n = (2n + 1)\pi/\beta$) and internal (particle-hole
and $b,a$) indices. Integrating over the Fermionic fields $\Psi$, one
obtains the ensemble averaged replicated partition function:
\begin{equation*}
  \langle \mathcal{Z}^n\rangle = \int D\Delta _1 D\Delta_2 \int
  D\Sigma e^{-\beta F}\; ,
\end{equation*}
where $F$ is the free energy of the system
\begin{multline}
  \beta F = \int_0^{\beta} d\tau \int d\mathbf{x} \left(2
  \frac{\Delta_1^2}{g_1} + 2 \frac{\Delta_2^2}{g_2}\right)\\ - \rm{tr}
  \ln \left(- \hat{\mathcal{G}}^{-1}\right) - \frac{\pi \nu_F}{\Gamma}
  \int d\mathbf{x} \, \rm{tr} \left[\Sigma(\mathbf{x}) \tau_3\right]^2
\label{eq:action}
\end{multline}
and $\hat{\mathcal{G}}$ is the quasi-particle matrix Green function in
the presence of disorder:
\begin{equation}
  - \hat{\mathcal{G}}^{-1} = - i\omega_n +
  \hat{\xi}_{\hat{\mathbf{p}}} \tau_{3} \sigma_{3} + \Delta_1
  \tau_1 + \Delta_2 \tau_{3}\sigma_1 + \Sigma (\mathbf{x})\; .
\label{eq:green}
\end{equation}
The matrix field $\Sigma (\mathbf{x})$ represents the contribution of
the non-magnetic impurity interaction to the self-energy.

The saddle-point associated with the action~\eqref{eq:action} obtained
by variation with respect to the self-energy $\Sigma$,
\begin{equation*}
  \Sigma (\mathbf{x}) = \frac{\Gamma}{2\pi \nu_F} \tau_3 \langle
  \mathbf{x}| \hat{\mathcal{G}} |\mathbf{x} \rangle \tau_3 \; ,
\end{equation*}
can be solved in the limit $\mu \gg \Gamma, \Delta_1, \Delta_2$, when
$\Sigma$, $\Delta_1$ and $\Delta_2$ can be considered homogeneous. In
this limit, which is compatible with the self-consistent Born
approximation, the Green function~\eqref{eq:green} is diagonal in
frequency and momentum space and can be explicitly inverted:
\begin{equation*}
  \mathcal{G}_{\omega_n,\mathbf{p}} = -\frac{i\tilde{\omega}_n +
  \xi_{\mathbf{p}} \tau_{3} \sigma_{3} + \tilde{\Delta}_1 \tau_1 +
  \tilde{\Delta}_2 \tau_{3} \sigma_1}{\tilde{\omega}_n^2 +
  \xi_{\mathbf{p}}^2 + \tilde{\Delta}_1^2 +
  \tilde{\Delta}_2^2}\; .
\end{equation*}
Here, we have defined the `renormalized' expressions for the frequency
and order parameters:
\begin{equation}
\begin{split}
  \displaystyle \tilde{\omega}_n \left(1 - \frac{\Gamma}{2}
  \frac{1}{\sqrt{\tilde{\omega}_n^2 + \tilde{\Delta}_1^2 +
  \tilde{\Delta}_2^2}} \right) &= \omega_n \\
  \tilde{\Delta}_1 \left(1 - \frac{\Gamma}{2}
  \frac{1}{\sqrt{\tilde{\omega}_n^2 + \tilde{\Delta}_1^2 +
  \tilde{\Delta}_2^2}} \right) &= \Delta_1\\
  \tilde{\Delta}_2 \left(1 + \frac{\Gamma}{2}
  \frac{1}{\sqrt{\tilde{\omega}_n^2 + \tilde{\Delta}_1^2 +
  \tilde{\Delta}_2^2}} \right) &= \Delta_2\; .
\label{eq:tilde1}
\end{split}
\end{equation}
From the above equations of motion, one may deduce that
\begin{equation}
  \frac{\tilde{\omega}_n}{\tilde{\Delta}_1} =
  \frac{\omega_n}{\Delta_1}\; ,
\label{eq:Delbar1}
\end{equation}
or, in other words, that in the weak disorder limit non magnetic
impurities do not suppress $s$-wave superconductivity (Anderson
theorem,\cite{anderson}) while, introducing the parameters $u =
\tilde{\omega}_n / \tilde{\Delta}_2$ and $\zeta = \Gamma /
\tilde{\Delta}_2$,
\begin{equation}
  \frac{\omega_n}{\Delta_2} = u \left[1 - \frac{\zeta}{\sqrt{1 + u^2
  \left(1 + \Delta_1^2 / \omega_n^2\right)}} \right] \; .
\label{eq:u}
\end{equation}

Finally, the self-consistency equations for the SC and EI order
parameters can be found minimizing the action~\eqref{eq:action} with
respect to $\Delta_{1,2}$:
\begin{equation}
  \Delta_{1,2} = \frac{\pi \lambda_{1,2}}{\beta} \sum_{\omega_n}
  \frac{\tilde{\Delta}_{1,2}}{\sqrt{\tilde{\omega}_n^2 +
  \tilde{\Delta}_1^2 + \tilde{\Delta}_2^2}} \; .
\label{eq:tilde2}
\end{equation}
Here, $\lambda_{1,2} = g_{1,2} \nu_{F}$ represent dimensionless
coupling constants. Note that, as in conventional BCS theory, the
integral over momentum can be performed by making use of the identity
$\int d\mathbf{p}/(2\pi)^3 = \int d\xi \nu(\xi) \simeq \nu_F \int
d\xi$. Employing Eq.~\eqref{eq:u}, the self-consistency equations can
then be rewritten in the form
\begin{equation}
\begin{split}
  \displaystyle \frac{1}{\lambda_1} &=\frac{\pi}{\beta}
  \sum_{\omega_n} \left[\Delta_1^2 + \omega_n^2
  \left(1+\frac{1}{u^2}\right)\right]^{-1 / 2} \\
  \displaystyle \Delta_2 &= \frac{\lambda_2 \pi}{\beta}
  \sum_{\omega_n} \left[1 + u^2 \left(1 +
  \frac{\Delta_1^2}{\omega_n^2}\right)\right]^{- 1/2}\; .
\end{split}
\label{eq:opars}
\end{equation}
Combining together equations~\eqref{eq:tilde1} with~\eqref{eq:opars},
we are now able to discuss the finite and zero temperature mean-field
phase diagram associated with the model~\eqref{eq:model}.

\section{Phase diagram}
\label{sec:phd}
%
\subsection{Temperature versus disorder phase diagram}
In the absence of disorder (i.e. $\Gamma =0$ and $u = \omega_n /
\Delta_2$), one may note that, except for different coupling
constants, the two self-consistency equations~\eqref{eq:opars} are
identical. Therefore, since they cannot be satisfied simultaneously,
even though the SC and EI \emph{instabilities} can occur
simultaneously, in clean materials the corresponding \emph{orderings}
are mutually exclusive. For $\lambda_1 > \lambda_2$ the system becomes
superconducting below
\begin{equation}
  \displaystyle T_1 (\Gamma = 0) = \frac{\gamma_E}{\pi} 2 \Omega_0 \,
  e^{-1/\lambda_1} \; ,
\label{eq:T1G0}
\end{equation}
where $\Omega_0$ is the frequency cutoff and $\gamma_E \simeq 1.78$ is
the Euler constant while, for $\lambda_2 > \lambda_1$, the transition
into the EI state occurs at
\begin{equation}
  \displaystyle T_2 (\Gamma =0) = \frac{\gamma_E}{\pi} 2 \Omega_0 \,
  e^{-1/\lambda_2} \; .
\label{eq:T2G0}
\end{equation}

Since charged non-magnetic impurities act as electron-hole pair
breaking perturbations, while they do not affect the SC state, for
$\lambda_1 > \lambda_2$ the SC state dominates at any disorder
strength $\Gamma$ and the EI state never appears. On the other hand,
for $\lambda_2 > \lambda_1$, the EI phase is energetically more
favorable at weak disorder, becomes suppressed for larger values of
$\Gamma$, and eventually gives way to superconductivity.  Non-magnetic
impurities suppress the EI state in exactly the same way as magnetic
impurities suppress the SC state.\cite{AbrikosovGorkov,Zittartz}
Therefore, one can infer that the dependence of the EI transition
temperature $T_2$ on the disorder strength $\Gamma$ is described by
the standard Abrikosov-Gor'kov expression:
\begin{equation}
  \ln \frac{T_2 (0)}{T_2 (\Gamma)} = \Psi \left(\frac{1}{2} +
  \frac{\Gamma}{2 \pi T_2 (\Gamma)}\right) - \Psi
  \left(\frac{1}{2}\right)\; .
\label{upperT2}
\end{equation}
At some critical disorder strength $\Gamma_{\ast}$, the EI and SC
transition temperatures eventually coincide $T_2 (\Gamma_{\ast}) = T_1
(\Gamma =0) = T_{\ast}$ and, for $\Gamma > \Gamma_{\ast}$, the system
becomes superconducting at the ($\Gamma $-independent) temperature
$T_1$ given by Eq.~\eqref{eq:T1G0}.

\begin{figure}
\begin{center}
\includegraphics[width=1\linewidth,angle=0]{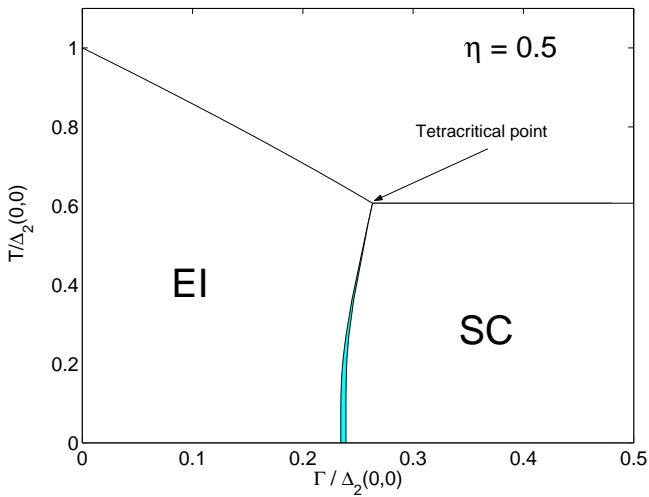}
\end{center}
\caption{\small The temperature vs disorder strength phase diagram of
        the model Eq.\eqref{eq:model} for $\protect\lambda
        _1^{-1}-\protect\lambda _2^{-1}=0.5$ . The dimensionless
        disorder strength is $\Gamma /\Delta _2(T=0,\Gamma =0)$ and
        temperature is measured in units of $T_{c2}(\Gamma =0)$. The
        EI and SC state are separated by a thin domain, in which the
        two order parameters coexist.}
\label{phdiag}
\end{figure}

One can therefore wonder in what way, at temperatures lower than
$T_{\ast}$, the transition between the EI and SC states takes
place. In Fig.~\ref{phdiag}, the temperature versus disorder phase
diagram is shown for values of the coupling constants such that $\eta
= \lambda_1^{-1} - \lambda_2^{-1} = 0.5$. The pure EI and SC states
are separated by a very thin region located in the $\Gamma <
\Gamma_{\ast}$ and $T < T_{\ast} $ region of the phase diagram, where
the two order parameters coexist. The three ordered phases (EI, SC,
and EI+SC) and the high-temperature disordered phase merge at the
tetracritical point $(\Gamma_{\ast} , T_{\ast})$. The boundaries
between the coexistence region and the two pure phases are critical
lines of second order transitions although, due to the very small
width of the coexistence region, the evolution of one pure phase into
another is close to being of first-order. This will be further
discussed in Sec.~\ref{sec:eisc}.

At the first critical line $T_1(\Gamma )$ separating the pure EI phase
from the mixed EI+SC phase, one has $\Delta_1 \to 0$ and $\Delta_2
\neq 0$. In the $\Delta_1 \to 0$ limit, the self-consistency
equations~\eqref{eq:opars} simplify to the expressions
\begin{equation}
\begin{split}
  \displaystyle \frac{1}{\lambda_1} &= 2\pi T_1 \sum_{\omega_n >0}
  \frac{u}{\omega_n \sqrt{1 + u^2}} \\
  \displaystyle \frac{\Delta_2}{\lambda_2} &= 2 \pi T_1
  \sum_{\omega_n >0} \frac{1}{\sqrt{1 + u^2}}\; .
\end{split}
\label{eq:Tc1}
\end{equation}
The second critical line $T_2 (\Gamma)$ separates the mixed state from
the pure SC state and is obtained by instead taking the limit
$\Delta_2 \to 0$ for nonzero $\Delta_1$. In this case, one can
approximate
\begin{equation*}
  u \simeq \frac{\omega_n}{\Delta_2} \left(1 +
  \frac{\Gamma}{\sqrt{\omega_n^2 + \Delta_1^2}}\right) \; ,
\end{equation*}
whereupon Eqs.~\eqref{eq:opars} take the form
\begin{equation}
\begin{split}
  \displaystyle \frac{1}{\lambda_1} &= 2 \pi T_2 \sum_{\omega_n
  >0} \frac{1}{\sqrt{\omega_n^2 + \Delta_1^2}} \\
  \displaystyle \frac{1}{\lambda_2} &= 2 \pi T_2 \sum_{\omega_n
  >0} \frac{1}{\sqrt{\omega_n^2 + \Delta_1^2} + \Gamma}\; .
\end{split}
\label{eq:Tc2}
\end{equation}
Note that the EI order parameter $\Delta_2$ appears at temperatures
lower than the ``upper'' $T_2 (\Gamma)$ given by Eq.\eqref{upperT2},
and disappears below the ``lower'' $T_2 (\Gamma)$, given by
Eq.~\eqref{eq:Tc2}.  Figure~\ref{phdiag} shows that $\Gamma_1$ (at
which the SC ordering sets in at $T=0$) and $\Gamma_2$ (at which the
EI ordering is destroyed at $T=0$) are smaller than the disorder
strength at the tetracritical point $\Gamma_{\ast}$. Therefore, in the
interval $\Gamma_2 < \Gamma < \Gamma_{\ast}$, the system passes
through three consecutive phase transitions as the temperature
decreases: Firstly the system becomes an excitonic insulator, then it
enters the mixed phase with the two coexisting order parameters, and
finally the growth of the SC order parameter with decreasing
temperature suppresses the EI ordering, resulting in the transition
into the pure SC state with $\Delta_2 = 0$.

\subsection{The zero temperature phase diagram}
\begin{figure}
\begin{center}
\includegraphics[width=1\linewidth,angle=0]{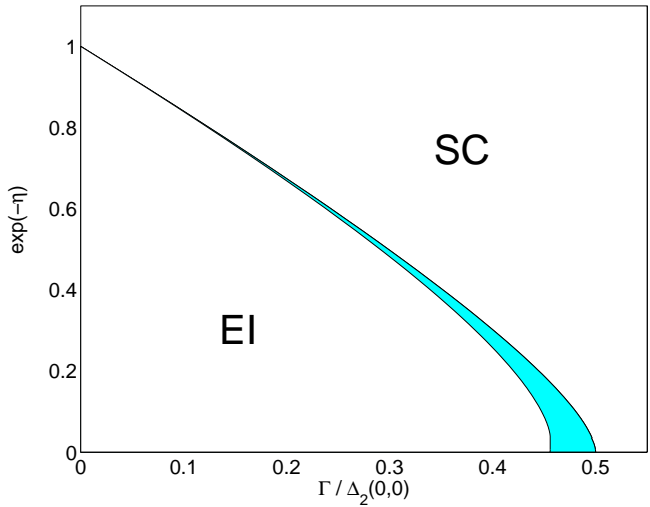}
\end{center}
\caption{\small The zero temperature phase diagram. Close to the
        quantum critical point $\eta = \Gamma = 0$, the coexistence
        region is extremely thin. For $\eta \ll 1$ it practically
        coincides with the disorder interval, in which the EI is
        gapless.}
\label{phdzt}
\end{figure}

The zero temperature phase diagram is shown in Fig.~\ref{phdzt}. The
EI state exists only for positive $\eta$ (or equivalently $e^{-\eta} =
\Delta_1 (T=0) / \Delta_2 (T=0 , \Gamma=0) < 1$). The coexistence
region (shaded) is confined between the two critical lines $\Gamma_1
(\eta) < \Gamma_2 (\eta)$. The system is superconducting for $\Gamma >
\Gamma_1$, while the excitonic condensate appears for $\Gamma <
\Gamma_2)$. For small $\eta$, i.e.  close to the quantum critical
point separating the EI and SC states in the absence of disorder, the
critical disorder values can be obtained as
\begin{equation}
\begin{array}{cl}
  \gamma_1 (\eta) &\simeq \frac{2}{\pi} \eta - \frac{3\pi^2 +
  8}{3\pi^3} \eta^2, \\ \\
  \gamma_2 (\eta) &\simeq \frac{2}{\pi} \eta - \frac{2\pi^2 -
  8}{\pi^3} \eta^2\; ,
\end{array}
\label{eq:ga1ga2}
\end{equation}
where $\gamma \equiv \Gamma / \Delta_2 (0,0)$. Therefore, the width of
the coexistence region is approximately given by $\gamma_2 - \gamma_1
\simeq 0.08 \eta^2$.

For large values of $\eta$, i.e. when the SC coupling $\lambda_1$ is
much smaller than the EI coupling $\lambda_2$, the coexistence region
essentially coincides with the disorder interval $e^{- \pi / 4} <
\gamma < 1 /2$ in which the EI state is
gapless.\cite{AbrikosovGorkov,Zittartz} Therefore, for $\eta > 1 +
3\pi / 4$, the superconductivity appears at the same disorder value,
at which the EI becomes gapless, $\gamma_1 = e^{- \pi / 4} \simeq
0.46$, while $\Gamma_2$ asymptotically approaches the disorder
strength at which the EI state is destroyed in the absence of
superconductivity:
\begin{equation}
  \gamma_2 \simeq 1 / 2 - \eta e^{-2\eta^2} \qquad \eta \gg 1\; .
\label{Gamma2etalarge}
\end{equation}
We note that, for $\eta < 1 + 3\pi / 4$, the gap in the spectrum of
quasi-particle excitations at zero temperature is nonzero for all
$\Gamma$, while for $\eta > 1 + 3\pi / 4$ it vanishes at a single
point $\Gamma = \Gamma_1$.

This re-entrant behavior and the form of the phase diagram are similar
to what was found for the spin-Peierls compound CuGeO$_{3}$ , which
upon doping shows an antiferromagnetic ordering coexisting with
spin-Peierls phase in some interval of doping
concentrations.\cite{MKK,Kiryukhin}

%

\section{EI-SC symmetry}
\label{sec:eisc}
\begin{figure}
\begin{center}
\includegraphics[width=1\linewidth,angle=0]{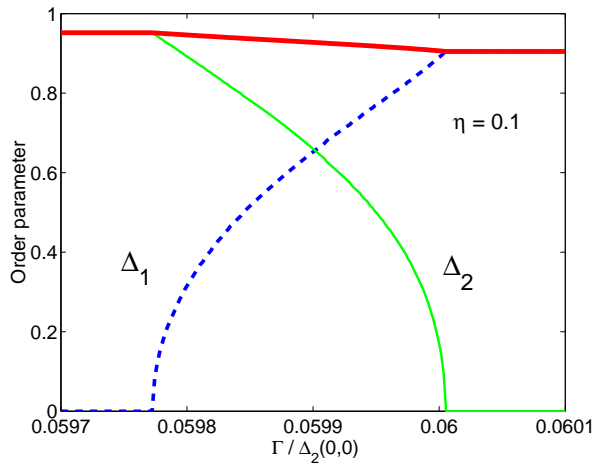}
\end{center}
\caption{\small The plots shows the dependence of the superconducting
        order parameter $\Delta_1$ (dashed line) and the excitonic
        order parameter $\Delta_2$ (thin solid line) on the disorder
        strength $\Gamma$ in the region of coexistence of the two
        phases for $\eta = 0.1$ and $T=0$. For
        $\Gamma_1<\Gamma<\Gamma_2$, $\Delta_1$ and $\Delta_2$ vary
        very fast, while $\Delta = \sqrt{{\Delta_1}^2 + {\Delta_2}^2}$
        (thick line) stays approximately constant.}
\label{fig:opars}
\end{figure}

To understand why the coexistence region is so narrow, it is
instructive to plot the EI and SC order parameters as functions of
$\Gamma$ in the coexistence region (see Fig.~\ref{fig:opars}). In the
interval $\Gamma_1 < \Gamma < \Gamma_2$ the excitonic(superconducting)
order parameter decreases (increases) fast with increasing $\Gamma$,
while $\Delta = \sqrt{\Delta_1^2 + \Delta_2^2}$ stays approximately
constant.

This behavior results from the symmetry between the $b$- and
$a$-electrons at the quantum critical point $\lambda_1 = \lambda_2$ in
the absence of disorder ($\Gamma =0$):
\begin{equation}
  \Psi \mapsto e^{i \phi \tau_2 \sigma_1/2} \Psi \; .
\label{eq:Psiphi}
\end{equation}
This transformation results is the rotation in the space of the two
order parameters over the angle $\phi \in [0, 2\pi]$,
\begin{align*}
  \Delta_1 &\mapsto \Delta_1 \cos \phi - \Delta_2 \sin \phi \\
  \Delta_2 &\mapsto \Delta_2 \cos \phi + \Delta_1 \sin \phi \; .
\end{align*}

At the mean-field level, the anomalous part of the average free energy
per unity of volume, $\langle f \rangle = \langle F \rangle/V$, can be
easily evaluated starting from~\eqref{eq:action} and making use of the
replica trick:
\begin{multline}
  \langle f \rangle = -\frac{4\pi}{\beta} \sum_{\omega_n}
  \left(\sqrt{\tilde{\omega}_n^2 + \tilde{\Delta}_1^2 +
  \tilde{\Delta}_2^2} - \left|\tilde{\omega}_n + \rm{sign}
  (\tilde{\omega}_n) \frac{\Gamma}{2}\right|\right) \\
  - \frac{2\pi \Gamma}{\beta} \sum_{\omega_n}
  \frac{\tilde{\Delta}_2^2}{\tilde{\omega}_n^2 +
  \tilde{\Delta}_1^2 + \tilde{\Delta}_2^2} +
  \frac{2}{\lambda_1} \Delta_1^2 + \frac{2}{\lambda_2}
  \Delta_2^2 \; .
\label{eq:avf1}
\end{multline}
Let us notice that, in the absence of disorder, the free energy only
depends on `the total gap' $\Delta = \sqrt{\Delta_1^2 +
\Delta_2^2}$. This follows from the fact that the generator of the
EI-SC rotations $\tau_2 \sigma_1$ commutes with the Hamiltonian
Eq.~\eqref{Hamiltonian} for $U(\mathbf{x})=0$. Moreover, for $g_1 =
g_2$, the last term in Eq.~\eqref{eq:avf1} is equal to $2 \Delta^2 /
g_1$ (the second term in Eq.~\eqref{eq:avf1} vanishes for $\Gamma
=0$). Thus, at the quantum critical point the free energy has a
`Mexican hat' profile as a function of the order parameters $(\Delta_1
, \Delta_2)$, symmetric under the rotations transforming the excitonic
insulator into the superconductor. This symmetry between
electron-electron and electron-hole pairing is analogous to the
symmetry unifying the $d$-wave superconductivity and
antiferromagnetism discussed in the context of high-T$_{c}$ and heavy
fermion materials.\cite{solomon,Zhang,Kitaoka}

Away from the quantum critical point, and for nonzero disorder, the
electron-hole symmetry is broken. Solving Eqs.~\eqref{eq:tilde1} for
$\tilde{\omega}_n $, $\tilde{\Delta}_1$, and $\tilde{\Delta}_2$,
perturbatively in the disorder strength $\Gamma $, and replacing at
$T=0$ the summations over the Matsubara frequency $\omega_n$ in
Eq.~\eqref{eq:avf1} by integrals, one can obtain an expansion of the
average energy density in powers of $\Gamma$:
\begin{multline}
  \frac{\langle f \rangle }{2\nu_{F}} = \frac{\Delta_1^2}{\lambda_1} +
  \frac{\Delta_2^2}{\lambda_2} - \Delta^2 \, \ln \left(2
  \frac{\Omega_0}{\Delta}\right) - \frac{1}{2} \Delta^2 \\ +
  \frac{\pi}{2} \frac{\Delta_2^2}{\Delta} \Gamma - \frac{1}{3}
  \frac{\Delta_2^2 (3 \Delta_1^2 + \Delta_2^2)}{\Delta^{4}} \Gamma^2
  \\ - \frac{\pi}{16} \frac{\Delta_1^2 \Delta_2^2 (\Delta_2^2 - 4
  \Delta_1^2)}{\Delta^{7}} \Gamma^{3} + O \left(\Gamma^{4}\right)\; .
\label{eq:enexp1}
\end{multline}

Denoting the dimensionless disorder strength by $\delta = \Gamma /
\Delta \ll 1$ and defining the angle $\phi$
\begin{align*}
  \Delta_1 &= \Delta \cos \phi & \Delta_2 &= \Delta \sin \phi \; ,
\end{align*}
we can recast Eq.~\eqref{eq:enexp1} in the form
\begin{multline}
  \frac{\langle f \rangle }{2\nu_{F}} \simeq \Delta^2 \left[\ln
  \frac{\Delta}{{\bar{\Delta}}} - \frac{1}{2} + \frac{\eta}{2} \cos
  2\phi + \frac{\pi\delta}{4} \left(1 - \cos 2\phi\right) \right. \\
  \left. - \frac{\delta^2}{12} \left(3 - 2\cos 2\phi - \cos
  4\phi\right) \right. \\ \left. + \frac{\pi \delta^{3}}{512} \left(6
  + 5 \cos 2\phi - 6\cos 4\phi - 5 \cos 6\phi\right) +
  O\left(\delta^{4}\right) \right] \; ,
\label{eq:enexp2}
\end{multline}
where ${\bar{\Delta}}$ is the geometric mean of the EI and SC order
parameters, $\bar{\Delta} = \sqrt{\Delta_1 \Delta_2 (0,0)} = 2\Omega_0
e^{- (\lambda_1^{-1} + \lambda_2^{-1})/2}$. The symmetry-breaking
terms in Eq.~\eqref{eq:enexp2} (that depend on the angle $\phi$) are
proportional to powers of $\eta$ and $\delta$. Thus, for $\eta ,
\delta \ll 1$, these terms are small and the energy has the slightly
deformed `Mexican hat' shape with an almost flat valley connecting the
points $\Delta_{\text{min}} (\phi)$, at which $\langle f \rangle$ has
a minimum for a given $\phi$.

\begin{figure}
\begin{center}
\includegraphics[width=1\linewidth,angle=0]{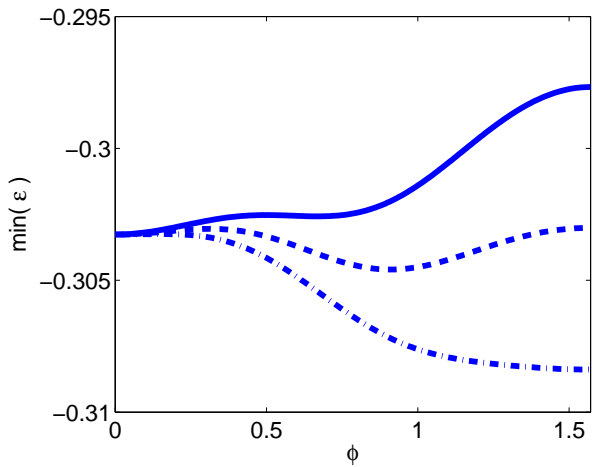}
\end{center}
\caption{\small The plot shows the energy minimum versus the angle
        $\phi $ at $\eta=0.5$ and for different values of $\Gamma$:
        $\Gamma =0.234<\Gamma _1$ (solid line), $\Gamma \sim
        \Gamma_{fo} = 0.237$ (dashed line) and $\Gamma =0.240>\Gamma
        _2$ (dot-dashed line).}
\label{fig:energ}
\end{figure}

%



This is illustrated in Fig.~\ref{fig:energ} where we plot the
$\phi$-dependence of the minimal energy density for, respectively,
$\Gamma = 0.234 < \Gamma_1$, $\Gamma = 0.237$ (in the coexistence
region) and $\Gamma = 0.240 > \Gamma_2$ (in both cases $\eta = 0.5
$). For $\Gamma < \Gamma_1$, the energy minimum is at $\phi = \pi / 2$
(the EI state), while at $\Gamma > \Gamma_2$ the energy minimum is for
$\phi =0 $ (the SC state). Though the $\phi$-dependence of the minimal
energy is, in general, rather complicated, the scale of the energy
variations in all three cases is very small, i.e. the valley is
practically flat. This is the reason for the narrow width of the
disorder interval $\Gamma_1 < \Gamma < \Gamma_2$, in which the two
phases coexist - a very small variation of the disorder strength
$\gamma$ is sufficient to shift the position the energy minimum from
$\phi = \pi / 2$ to $\phi = 0$ along the energy valley, in which
$\Delta = \sqrt{\Delta_1^2 + \Delta_2^2}$ remains practically
unchanged.

Figure~\ref{fig:energ} also illustrates the absence of first-order
transitions in the model\eqref{eq:model}. Note that $\Gamma = 0.237$
is close to $\Gamma_{\text{fo}}$, at which the energies of the SC and
EI states become equal: $ \langle f (0)\rangle = \langle f(\pi /
2)\rangle$. The first-order transition between the two pure states,
however, does not occur, since the energy has the global minimum at
some angle $\phi$, such that $0 < \phi < \pi / 2$, corresponding to a
mixed ground state.  Furthermore, when $\Gamma = 0.237$, the energy
has a local maximum at $\phi = \pi / 2$, enforcing the EI state to be
metastable.


\section{Discussion and Conclusions}
\label{sec:conclusions}

We discussed effects of disorder in systems with competing
instabilities, such as CDW superconductors. We considered a simple
model, which describes a metal with two perfectly nested electron-like
and hole-like parts of the Fermi surface. In this model the interplay
between the electron-electron and electron-hole pairings is very
strong, as these they compete over the whole Fermi surface.

We showed that disorder can be used to tune the balance between the
two competing phases and to stabilize the state, in which they
coexist. The charged nonmagnetic impurities induce superconductivity
by suppressing the CDW state. Such a disorder-induced
superconductivity is observed in the irradiated two-dimensional CDW
material $2H$-TaS$_{2}$.\cite{mutka} In other transition metal
dichalcogenides, e.g. $2H$-NbSe$_{2}$ and $2H$-TaSe$_{2}$, which are
CDW superconductors already in absence of disorder, a small amount of
irradiation-induced defects results in an enhancement of
$T_{c}$.\cite{mutka} Similar behavior is observed in the
quasi-one-dimensional CDW material Nb$_{1-x}$Ta$_{x}$Se$_{3}$. In the
pure NbSe$_{3}$, $T_{c}$ is smaller than $50$ mK at ambient
pressure.\cite{Monceau} The substitution of Nb for Ta suppresses the
resistivity anomalies due to the CDW transitions, while $T_{c}$ grows
up to $\sim 2$K at $x=0.05$.\cite{Fuller} The effect of impurities in
these materials is similar to that of pressure and hydrogen
intercalation.\cite{Monceau,Berthier,Murthy}

In agreement with these experimental findings, the phase diagrams of
our model Figs.~\ref{phdiag} and \ref{phdzt}, show a strong
sensitivity of the ground state to disorder and the coexistence of the
SC and EI states in the presence of disorder. This behavior can be
easily understood and described analytically, using the Landau
expansion of the free energy in powers of the EI and SC order
parameters near the quantum critical point [see Eq.(\ref{eq:enexp1})],
which we derived from the microscopic model. Disorder distorts the
shape of the energy potential and continuously shifts the position of
the minimum from the point corresponding to the excitonic insulator to
the point corresponding to the superconducting state, which gives rise
to the coexistence of the two states.

The microscopic origin of this coexistence is the break-up of a part
of the electron-hole pairs by disorder and the subsequent
recombination of the released fermions into electron-electron and
hole-hole pairs. In other words, disorder transforms the CDW gap in
the single-electron density of states into a pseudogap, filled with
states describing the broken electron-hole pairs. The SC phase
develops inside this pseudogap, which resembles the behavior observed
in high-T$_{c}$ cuprates.\cite{Timusk}

In addition to the disorder-induced superconductivity, resulting from
the suppression of the EI state, the phase diagram of our model (see
Fig.~\ref{phdiag}) shows an interesting `inverse' effect, namely the
suppression of the EI state due to the growth of the SC order
parameter with decreasing temperature. Though this re-entrance
transition is just another consequence of the competition between the
two types of ordering, we did not find any reports of such a behavior
in CDW superconductors in the literature. This, however, it has been
observed in the quasi-one-dimensional spin-Peierls compound
CuGeO$_{3}$, where impurities induce the long range N\'{e}el
ordering.\cite{Kiryukhin} In this material, the interplay between the
dimerized and antiferromagnetic states allows for a similar
theoretical description.\cite{MKK} In most CDW superconductors the CDW
transition occurs at a much higher temperature than the SC transition,
so that the influence of the Cooper pairing on the CDW modulations is
difficult to observe. Furthermore, the CDW gap only opens on a nested
part of the Fermi surface. In quasi-one-dimensional NbSe$_{3}$ the
fraction of the Fermi surface affected by the CDW transition was
estimated to be $\sim 0.6$ at ambient pressure.\cite{Fuller} In the
two-dimensional $2H$-NbSe$_{2}$ this fraction is apparently very
small, since the gap opening actually increases the conductivity of
this material\cite{harper} and the part of the Fermi surface, where
the gap opens, was not found in ARPES
experiments,\cite{straub,Valla,Yokoya} even though the gap value
(34meV) is known.\cite{Wang} For partially gapped Fermi surfaces the
competition between the CDW and SC states is less strong, so that in
$2H$-NbSe$_{2}$ they coexist even in absence of disorder.

While the enhancement of the SC transition temperature upon the
suppression of the CDW state is well documented in many materials, the
experimental situation with influence of the superconductivity on the
CDW state is less clear. On the one hand, Raman experiments on
$2H$-NbSe$_{2}$ show the suppression of the intensity the collective
SC mode by magnetic field with the concomitant enhancement of the
intensity of the CDW modes.\cite{Klein,Peter} On the other hand, no
effect of the superconducting ordering below $7$K and of the
suppression of the SC state by magnetic field on the CDW modulation
was observed in x-ray experiments.\cite{Du} The understanding of the
behavior of $2H$-NbSe$_{2}$ is complicated by the multi-sheet
structure of the Fermi surface and the momentum- and sheet-dependence
of both order parameters. \cite{Yokoya} The interplay between the CDW
and SC states in this and other materials requires further
experimental and theoretical studies.

Crucially, one may note that the phase diagram of the interacting
system was inferred from the self-consistent Hartree-Fock
approximation which captures only the mean-field characteristics. In
view of the filamentary structure of the coexistence region, the
system can be susceptible to mesoscopic or sample to sample
fluctuations due to the quenched impurity potential. Such effects are
recorded in fluctuations of the field $\Sigma (\mathbf{x})$ around its
saddle-point or mean-field value (as opposed to the leading terms
gathered in the low-order $\Gamma$ expansion considered here). In the
gapless regime, such effects can give rise to long-ranged diffusion
mode contributions to the generalized pair susceptibility (cf., e.g.,
Ref.~[\onlinecite{spivak_zhou}]). However, in the present case, the
disorder potential imposes a symmetry breaking perturbation on the EI
phase. As such, we can expect mesoscopic fluctuations due to disorder
to impose only a short-ranged (i.e. local on the scale of the
coherence length of the EI order parameter) perturbation on the pair
susceptibility. In the vicinity of the coexistence region, where the
potential for the angle $\phi$ is shallow, the effect of these
mesoscopic fluctuations may be significant.

To understand the effect of random fluctuations in the coexistence
region, we consider the Ginzburg-Landau expansion for the ground state
energy close to the quantum critical point $\eta=0,\Gamma=0$, at which
the EI and SC states are degenerate. In the vicinity of this point the
phase $\phi$ of the `total' order parameter $\Delta_1+i\Delta_2 =
\Delta e^{i\phi}$ is a soft mode, so that weak disorder mainly induces
spatial fluctuations of the phase, while the magnitude of the order
parameter $\Delta$ approximately stays constant. As in the derivation
of Eq.(\ref{eq:enexp2}), we expand the energy in powers of $\Delta$
and disorder strength, assuming that $\Gamma \ll \Delta$, which is
justified in the coexistence region, where $\Gamma \sim
\frac{2\eta}{\pi}\Delta$ (see Eq.(\ref{eq:ga1ga2}). Assuming that the
phase varies slowly on the length scale of the correlation length $\xi
= \frac{v_F}{\Delta}$, where $v_F$ is the Fermi velocity, we obtain
\begin{equation}
  F \simeq \int d \mathbf{x}\left[\nu_F \left( \frac{v_F^2}{6} (\nabla
  \phi)^2+u(\mathbf{x}) \cos2\phi\right)+ \langle f \rangle \right]\;
  ,
\end{equation}
where the first term describes the `elastic energy' of an
inhomogeneous state, the disorder-averaged free energy $\langle f
\rangle$ is given by Eq.(\ref{eq:enexp2}), and $u(\mathbf{x})$ is the
fluctuating part of disorder coupled to the phase of the order
parameter. Neglecting correlations on a scale smaller that the
correlation length, $u(\mathbf{x})$ can be approximately considered as
a random $\delta$-correlated Gaussian variable with zero average,
$\langle u(\mathbf{x}) \rangle = 0$, and variance
\begin{align}
  \langle u({\bf x}) u({\bf y}) \rangle &= A \delta({\bf x}-{\bf y}) &
  A &= \frac{\pi^4\Gamma^2\Delta^2}{2k_F^3} \; ,
\end{align}
(we omit the lengthy calculations that lead to this result). The
coupling to disorder also occurs in higher orders of the expansion,
but those terms are relatively small and can be neglected.

Following the Imry and Ma argument,\cite{ImryMa} we consider a large
phase fluctuation, e.g., a droplet of the SC phase of the spatial
extent $L$ inside the EI matrix.  Comparing the typical energy gain
due to the coupling to disorder $\sim \nu_F\sqrt{AL^3}$ with the loss
in the elastic energy $\sim \nu_F v_F^2 L$, we find that the
fluctuation is energetically favorable for
\begin{equation}
  L > \frac{v_F^4}{A} = \frac{v_F^4 k_F^3}{2\pi^2 \eta^2\Delta^4}\; ,
\label{eq:L1}
\end{equation}
where we took into account that, in the coexistence region, $\Gamma
\sim \frac{2\eta}{\pi}\Delta$.

The crucial difference of our model from that consi\-de\-red in
Ref.[\onlinecite{ImryMa}] is the absence of an exact continuous
symmetry. Even in the coexistence region, the minimal-energy valley
connecting the SC and EI points ($\phi = 0$ and $\phi =
\frac{\pi}{2}$) is not perfectly flat. The typical amplitude of the
variations of the energy density is $\sim \eta^2 \nu_F \Delta^2$ (see
Eq.~\eqref{eq:enexp2}), resulting in the energy loss $\sim \eta^2
\nu_F \Delta^2 L^3$ proportional the volume of the fluctuation, which
suppresses large droplets. Comparing it with the energy gain, we find
\begin{equation}
  L^3 < \frac{A}{\eta^4\Delta^4}\; . 
\label{eq:L2}
\end{equation}
Equations (\ref{eq:L1}) and (\ref{eq:L2}) hold simultaneously for
\begin{equation}
v_Fk_F < \eta \Delta\; , 
\label{eq:condition}
\end{equation}
which cannot be satisfied in the weak coupling limit. One may wonder
why the condition (\ref{eq:condition}) does not hold even for $\eta =
\Gamma = 0$, where the model has a continuous symmetry. The reason is
that in our model the role of disorder is two-fold. On the one hand,
it couples to the order parameter, as in the `random field' model
discussed in Ref.[\onlinecite{ImryMa}] and tends to destroy the
ordering. On the other hand, it affects the energy difference between
the EI and SC states and, therefore, suppresses the phase
fluctuations, by destroying the symmetry of the energy potential. The
second effect, which is linear in $\Gamma$, is stronger than the
first.

Thus, the inhomogeneity of the order parameter, resulting from typical
disorder fluctuations is small. The phase fluctuations can also be
induced by large disorder fluctuations ('Lifshitz tails'), but their
contribution to the free energy is exponentially small.\cite{Dotsenko}
This justifies our mean field treatment of disorder.

This conclusion may not hold, however, for strongly coupled CDW
superconductors or for other types of disorder. Qualitatively, we
expect that inhomogeneous excitonic and superconducting order
parameters may result in a broadening of the coexistence region. The
local suppression of the excitonic pairing near charged impurities can
give rise to the local enhancement of the superconducting order. The
state with such a nanoscale phase separation, in which two competing
orders alternate in antiphase without a loss of the macroscopic
coherence, can be more energetically favorable than the uniform state
and, therefore, can be stabilized in a wider interval of
parameters. Such a state was observed in $\mu$SR experiments on doped
CuGeO$_3$, which shows both spin-Peierls and antiferromagnetic
ordering.\cite{kojima}

In conclusion, we studied effects of disorder on systems with
competing superconducting and charge-density-wave instabilities.  We
showed that even in the extreme situation, when the competition takes
place over the whole Fermi surface and the superconducting and
charge-density-wave phases are mutually exclusive, disorder can give
rise to their coexistence in a spatially homogeneous
state. Furthermore, disorder itself can be used as a parameter, with
which one can tune the balance between competing phases.  Although our
model is too simple to describe the physics behind the coexistence of
superconductivity and CDW(SDW) states in, e.g., high-$T_c$ or heavy
fermion materials, we believe that the ability of disorder to bring
together incompatible phases may be important for understanding phase
diagrams of these systems.

\appendix

\section{Coexisting instabilities and derivation of the effective model}
\label{AppendixA}
In this appendix we obtain a condition under which both the SC and EI
instabilities can occur simultaneously. Here, we consider more
realistic interactions between electrons than those described by the
model~\eqref{eq:model}, namely, the phonon-mediated interaction and
the Coulomb repulsion. The Coulomb repulsion counteracts the
phonon-mediated attraction between electrons and suppresses the SC
instability. The same holds for the instability towards the formation
of the excitonic condensate with the difference that the two types of
interaction now change roles: the Coulomb force favors the
electron-hole pairing, while the one-phonon exchange results in a
repulsion between electrons and holes. We will show that the SC and EI
instabilities can coexist due to different frequency dependence of the
two types of interactions.

For retarded phonon-mediated interactions, the order parameters
$\Delta_1$ and $\Delta_2$ are frequency-dependent, which complicates
the solution of the self-consistency equations. We show, however, that
in the weak coupling and weak disorder limit, the equations for the
order parameters at zero frequency coincide with
Eq.~\eqref{eq:tilde2}, which justifies the model introduced in
Sec.~\ref{sec:model}. Moreover, we will give the explicit expressions
for the coupling constants $g_1$ and $g_2$ appearing in
Eq.~\eqref{eq:model}.

We describe effective electron-electron interactions by a non-local
action
\begin{multline}
  S_{\text{eff}} = \frac{g_{\text{e-ph}}^2}{2}\int d \mathbf{x} \int
  d\tau d\tau' \rho (\mathbf{x},\tau) D (\tau -\tau') \rho (\mathbf{x}
  , \tau') \\
  + \frac{g_C^2}{2} \int d\mathbf{x} d\tau \rho^2 (\mathbf{x},\tau) \;
  ,
\label{Seff}
\end{multline}
where $\rho = \sum_{\sigma} (\psi_{a\sigma}^{\dag} \psi_{a\sigma} +
\psi_{b\sigma}^{\dag} \psi_{b\sigma}) $ is the total electron density.
The first term is the phonon-mediated effective attraction between
electrons and $D (\tau - \tau')$ is the phonon Green function. For a
single dispersionless optical phonon with the frequency $\Omega_0$ and
the propagator $D_{\omega_n} = - \Omega_0^2/ (\omega_n^2 +
\Omega_0^2)$, we have $D (\tau -\tau') = - \Omega_0 e^{-\Omega_0 |\tau
-\tau'|}/2$ for $T \ll \Omega_0$. The second term in Eq.~\eqref{Seff}
is the instantaneous Coulomb interaction. We neglect the momentum
dependence of the screened electron-phonon and Coulomb couplings,
which makes the electron-electron interactions local in space.

The couplings for the $a$ and $b$-electrons in Eq.(\ref{Seff}) give
rise to a large freedom in the choice of order parameters, which in
reality may not be present, e.g., due to the inter-band scattering,
which separately does not conserve the numbers of the $a-$ and
$b-$electrons. In what follows we restrict ourselves to the anomalous
averages considered in Sec.~\ref{sec:model}, which, for retarded
interactions Eq.~\eqref{Seff}, are time-dependent ($j = a,b$ and
$\sigma = \uparrow ,\downarrow$):
\begin{align*}
  \Delta_1 (\tau -\tau') &= -\left[g_{\text{e-ph}}^2 D (\tau -\tau') +
  g_C^2\right] \langle \psi_{i\uparrow} (\tau) \psi_{i\downarrow}
  (\tau') \rangle\\
  \Delta_2 (\tau -\tau') &= -\left[g_{\text{e-ph}}^2 D (\tau -\tau') +
  g_C^2\right] \langle \psi_{a\sigma} (\tau) \psi_{b\sigma} (\tau')
  \rangle\; .
\end{align*}
In the frequency representation the self-consistency equations read
\begin{align*}
  \Delta_1 (\omega_n) &= -\frac{T}{4 V} \sum_{\mathbf{p} \omega_n'}
  \left(g_{\text{e-ph}}^2 D_{\omega_n -\omega_n'} - g_C^2\right)
  \rm{tr} \left(\tau_1 G_{\mathbf{p}, \omega_n'}\right) \\
  \Delta_2 (\omega_n) &= \frac{T}{4 V} \sum_{\mathbf{p} \omega_n'}
  \left(g_{\text{e-ph}}^2 D_{\omega_n - \omega_n'} - g_C^2\right)
  \rm{tr} \left(\tau_{3} \sigma_1 G_{\mathbf{p} , \omega_n'}\right) \;
  ,
\end{align*}
where the electron Green function is given by Eq.~\eqref{eq:green}.

To simplify the algebra, we consider here only the zero temperature
case. The integration over the electron excitation energy $\xi$ gives
\begin{widetext}
\begin{equation}
\begin{split}
  \Delta_1 (\omega) = \frac{1}{2} \int_{-E_{c}}^{+E_{c}}d \omega'
  \left[\kappa_1 \frac{\Omega_0^2}{(\omega -\omega')^2 + \Omega_0^2} -
  \kappa_2\right] \frac{\tilde{\Delta}_1
  (\omega')}{\sqrt{\tilde{\omega}'^2 + \tilde{\Delta}_1^2 (\omega') +
  \tilde{\Delta}_2^2 (\omega')}} \\
  \Delta_1 (\omega) = \frac{1}{2} \int_{-E_{c}}^{+E_{c}}d \omega'
  \left[\kappa_2 - \kappa_1 \frac{\Omega_0^2}{(\omega -\omega')^2 +
  \Omega_0^2} \right] \frac{\tilde{\Delta}_1
  (\omega')}{\sqrt{\tilde{\omega}'^2 + \tilde{\Delta}_1^2 (\omega') +
  \tilde{\Delta}_2^2 (\omega')}}
\end{split}
\label{Delta12omega}
\end{equation}
\end{widetext}
where we have introduced the dimensionless coupling constants,
$\kappa_1=\nu _{F} g_{\text{e-ph}}^2$ and $\kappa_2=\nu_{F} g_C^2$,
and where $E_{c}$ is the frequency cutoff required for the
instantaneous Coulomb interaction. Moreover the variables
$\tilde{\omega},\tilde{\Delta}_1,$ and $\tilde{\Delta}_2$ are defined
in~\eqref{eq:tilde1}.

Although Eqs.~\eqref{Delta12omega} look at a first sight complicated,
one can see that, in the limit of weak coupling and weak disorder,
$\Delta_{1,2} , \Gamma \ll \Omega_0 <E_c$, their solution can be found
by making use of the fact that the order parameters $\Delta_1 (\omega)
$ and $\Delta_2 (\omega)$ strongly vary at frequencies $\omega \sim
\Omega_0$, while $\tilde{\omega}$, $\tilde{\Delta}_1$, and
$\tilde{\Delta}_2$ are nontrivial functions of $\omega$ only at much
lower frequencies $\omega \sim \Gamma$, where $\Delta_1 (\omega)$ and
$\Delta_2 (\omega)$ can be replaced by their zero frequency
values. Therefore, we can solve Eqs.~\eqref{Delta12omega} in two
steps: first we find the frequency dependence of the order parameters
$\Delta_1(\omega )$ and $\Delta_2(\omega) $ for arbitrary values of
$\Delta_1(0)$ and $\Delta_2 (0)$, and then we solve the
self-consistency equations for $\Delta_1 (0) $ and $\Delta_2 (0)$.

It is convenient to use the dimensionless variables $x = \omega /
\Omega_0$ and $y_{1,2} (x) = \Delta_{1,2} (\omega) / \Omega_0$ (and
similarly $\tilde{x}$ and $\tilde{y}_{1,2} (x)$), in terms of which
the first of the equations~\eqref{Delta12omega} reads
\begin{multline*}
  y_1 (x) = \int_0^{\Lambda} dx' \frac{\tilde{y}_1
  (x')}{\sqrt{\tilde{x}'^2 + \tilde{y}_1^2 (x') + \tilde{y}_2^2 (x')}}
  \\
  \left\{\frac{\kappa _1}{2} \left[\frac{1}{(x+x')^2 + 1} +
  \frac{1}{(x-x')^2 + 1}\right] -\kappa_2\right\}\; ,
\end{multline*}
where $\Lambda = E_{c} / \Omega_0$. We then introduce an intermediate
scale $X$, such that $y_{1,2}\ll X\ll 1$. In the interval $0 \leq x'
\leq X$, we can neglect the $x'$-dependence of the kernel of this
integral equation and the functions $y_{1,2} (x')$ (however,
$\tilde{y}_{1,2}$ and $\tilde{x}'$ still do depend on $x'$). In the
second interval $X \leq x' \leq \Lambda $, we substitute $ \tilde{y}_1
/\sqrt{\tilde{x}'^2 + \tilde{y}_1^2 + \tilde{y}_2^2}$ by $y_1 (x') /
x'$ and perform the integration by parts. In this way we obtain
\begin{widetext}
\begin{multline}
  y_1 (x) = - y_1 (\Lambda) \kappa_2 \ln \Lambda +
  \left(\frac{\kappa_1}{1 + x^2} - \kappa_2\right)
  \left[\int_0^X dx' \frac{\tilde{y}_1 (x')}{\sqrt{\tilde{x}'^2
  + \tilde{y}_1^2 (x') + \tilde{y}_2^2 (x')}} - y_1 (0) \ln X \right]\\
  - \int_0^{\infty} dx' \ln x' \frac{d}{dx'}
  \left\{\left[\frac{\kappa_1}{2} \left( \frac{1}{(x+x')^2 + 1} +
  \frac{1}{(x-x')^2 + 1}\right) - \kappa_2\right] y_1 (x') \right\}\;
  ,
\label{eq:y1}
\end{multline}
\end{widetext}
where the limits of the second integration were extended to $0$ and
$\infty$, as there is convergence both at small and large frequencies.

Since, at $X \gg y_{1,2}$, Eq.~\eqref{eq:y1} is independent of $X$, we
can chose $X=1$ (and still substitute $y_{1,2} (x')$ by $y_{1,2} (0)$
in the first integral). The value of $y_1$ at the cutoff is then given
by
\begin{multline*}
  y_1 (\Lambda) = - \kappa _2^* \left[\int_0^1 dx' \frac{\tilde{y}_1
  (x')}{\sqrt{\tilde{x}'^2 + \tilde{y}_1^2 (x') + \tilde{y}_2^2
  (x')}}\right. \\
  \left. - \int_0^{\infty} dx \ln x \frac{dy_1}{dx} \right] \; ,
\end{multline*}
where $\kappa_2^{\ast} = \kappa_2/ (1+\kappa_2 \ln \Lambda)$. For
arbitrary $x$ we have
\begin{multline}
  y_1 (x) = \left(\frac{\kappa_1}{1+x^2} - \kappa_2^{\ast}\right) I
  (y_1 (0) , y_2 (0))\\
  - \int_0^{\infty} dx' \ln x' \frac{d}{dx'} \left\{
  \left[\frac{\kappa_1}{2} \left(\frac{1}{(x+x')^2 + 1}\right.\right.\right. \\
  \left. \left.\left. +\frac{1}{ (x-x')^2 + 1}\right)
  -\kappa_2^{\ast}\right] y_1 (x')\right\}\; ,
\label{eq:y1y1}
\end{multline}
where the notation
\begin{equation*}
  I (y_1 (0) , y_2 (0)) = \int_0^1 dx' \frac{\tilde{y}_1
  (x')}{\sqrt{\tilde{x}'^2 + \tilde{y}_1^2 (x') + \tilde{y}_2^2 (x')}}
\end{equation*}
is used to stress the fact that $y_1$ and $y_2$ are assumed to be
frequency independent.

In the weak coupling limit the first term in the right-hand side of
Eq.~\eqref{eq:y1y1}, proportional to the `large logarithm' $\ln
\sqrt{y_1^2 + y_2^2}$, is much larger than the second term, so this
integral equation can be solved by iterations, which generate a
perturbative expansion for $y_1 (x)$. To the lowest order, the
frequency dependence of the order parameter coincides with that of the
kernel:\cite{AndersonMorel}
\begin{equation}
  y_1 (x) = \left(\frac{\kappa_1}{1+x^2} - \kappa_2^{\ast}\right) I
  (y_1 (0) , y_2 (0))\; .
\label{eq:y1(0)}
\end{equation}
Then the self-consistency equation for $\Delta_1 (0)$ coincides with
Eq.~\eqref{eq:tilde2} at $T=0$:
\begin{equation}
  \Delta_1 (0) = \lambda _1 \int_0^{\Omega_0} d\omega
  \frac{\tilde{\Delta}_1 (\omega)}{\sqrt{\tilde{\omega}^2 +
  \tilde{\Delta}_1^2 (\omega) + \tilde{\Delta}_2^2 (\omega)}} \; ,
\label{eq:Delta1(0)}
\end{equation}
and the effective coupling constant is given by
\begin{equation}
  \lambda_1 = \kappa_1 - \kappa_2^{\ast} = \kappa_1 -
  \frac{\kappa_2}{1 + \kappa_2 \ln E_{c} / \Omega _0}\; .
\label{eq:lambda1(0)}
\end{equation}
The negative term in the coupling constant describes the reduction of
the attraction between electrons due to the Coulomb repulsion, but
this reduction is itself reduced by the presence of the large
logarithm in the denominator due to the difference in the time scales
of the retarded phonon-mediated attraction and the Coulomb
repulsion.\cite{Bogoliubov,AndersonMorel} The first-order correction
to $y_1 (x)$, found by substituting Eq.~\eqref{eq:y1(0)} into the
integral in Eq.~\eqref{eq:y1y1} (as well as all higher-order
corrections), leaves the form of the self-consistency
equation~\eqref{eq:Delta1(0)} unchanged, but results in a small
modification of the expression for the effective coupling constant
through $\kappa_1$ and $\kappa_2$:
\begin{equation*}
  \lambda_1 = \kappa_1 \left(1 - \frac{\kappa_1}{2}\right) -
  \kappa_2^{\ast }\; .
\end{equation*}

The frequency dependence of the excitonic insulator order parameter
$\Delta_2 (\omega) $ and the self-consistency equation for $\Delta_2
(0)$ can be obtained from
Eqs.(\ref{eq:Delta1(0)},\ref{eq:lambda1(0)}) by
the substitution $\Delta_1 \mapsto \Delta_2$, $\kappa_1 \mapsto -
\kappa_1$, and $\kappa_2 \mapsto - \kappa_2$:
\begin{equation}
  \Delta_2 (0) = \lambda_2 \int_0^{\Omega_0} d\omega
  \frac{\tilde{\Delta}_2 (\omega)}{\sqrt{\tilde{\omega}^2 +
  \tilde{\Delta}_1^2 (\omega) + \tilde{\Delta}_2^2 (\omega)}} \; ,
\label{eq:Delta2(0)}
\end{equation}
where the effective coupling constant $\lambda_2$ is to the lowest
order given by
\begin{equation}
  \lambda_2 = \frac{\kappa_2}{1 - \kappa_2 \ln E_{c} / \Omega_0} -
  \kappa_1 \; .
\label{eq:lambda2(0)}
\end{equation}

In Fig.~\ref{fig:opfreq} we show the typical frequency dependence of
the SC and EI order parameters, calculated for $\kappa_1 = 0.25$ and
$\kappa_2 = 0.1$ and $\Gamma=0$.  (Since in the absence of disorder the SC
and EI states cannot coexist, we calculated $\Delta_1 (\omega)$
assuming $\Delta_2 = 0$ and vice versa.) The SC order parameter
$\Delta_1 (\omega)$ is positive at small frequencies and changes sign
at $\omega \simeq \Omega_0$, while the EI order parameter $\Delta_2
(\omega)$ has a dip for $|\omega| < \Omega_0$. The `separation' of the
two order parameters in frequency is crucial for the coexistence of
instabilities.

The necessary condition for superconductivity to appear is $\lambda_1
> 0$, while the instability towards the excitonic condensate occurs
for $\lambda_2>0$. These two conditions,
\begin{align*}
  \lambda_1 &> \frac{\lambda_2}{1 + \lambda_2 \ln \Lambda} &
  \lambda_1 &< \frac{\lambda_2}{1 - \lambda_2 \ln \Lambda} \; ,
\end{align*}
hold simultaneously for
\begin{equation}
  |\eta| = \left|\frac{1}{\lambda_1} - \frac{1}{\lambda_2}\right| <
   \ln \frac{E_{c}}{\Omega _0} \; .
\end{equation}

A weak disorder has little effect on the frequency dependence of
$\Delta_1$ and $\Delta_2$. However, its presence is crucial for the
stabilization of the mixed state, in which the two order parameters
coexist.

\begin{acknowledgments}
  We are indebted to D. Khmel'nitskii and D. Khomskii for valuable
  discussions. This work was supported by the MSC$^{\mathit{plus}}$
  program. The financial support of the British Council and the
  Trinity college is gratefully acknowledged. One of us (FMM) would
  like to acknowledge the financial support of EPSRC (GR/R95951).
\end{acknowledgments}


\end{document}